# Angular Independent Photonic Pigments via the Controlled Micellization of Amphiphilic Bottlebrush Block Copolymers

*Tianheng H. Zhao,[1] Gianni Jacucci,[1] Xi Chen,[2] Dong-Po Song,[2] \* Silvia Vignolini[1] \* and Richard M. Parker[1] \**

Tianheng H. Zhao, Gianni Jacucci, Prof. Silvia Vignolini, Dr. Richard M. Parker
Department of Chemistry, University of Cambridge, Lensfield Road, Cambridge CB2 1EW, UK.
E-mail: sv319@cam.ac.uk, rmp53@cam.ac.uk

Xi Chen, Dr. Dong-Po Song
Tianjin Key Laboratory of Composite and Functional Materials, School of Materials Science and Engineering, Tianjin University, Tianjin 300350, China.
E-mail: dongpo.song@tju.edu.cn

## Abstract

Photonic materials with angular independent structural color are highly desirable because they offer the broad viewing angles required for application as colorants in paints, cosmetics, textiles or displays. However, they are challenging to fabricate as they require isotropic nanoscale architectures with only short-range correlation. In this article, porous microparticles with such a structure are produced in a single, scalable step from an amphiphilic bottlebrush block copolymer. This is achieved by exploiting a novel 'controlled micellization' self-assembly mechanism within an emulsified toluene-in-water droplet. By restricting water permeation through the droplet interface, the size of the pores can be precisely addressed, resulting in structurally colored pigments. Furthermore, the reflected color can be tuned to reflect across the full visible spectrum using only a single polymer ($M_n$ = 290 kDa) by altering the initial emulsification conditions. Such 'photonic pigments' have several key advantages over their crystalline analogues, as they provide isotropic structural coloration that suppresses iridescence and improves color purity without the need for either refractive index matching or the inclusion of a broadband absorber.

Synthetic photonic materials typically exploit photonic crystal structures, where the scattering elements are periodic over long distances, with common examples ranging from the top-down deposition of alternating dielectrics to produce a 1-D multi-layer, to the self-assembly of colloidal particles to produce 3-D colloidal crystals and opals.[1,2] However, there is a growing interest in structures where only short-range order is present. Such isotropic structures, often referred to as photonic glasses, can be uniquely used to obtain non-iridescent structural coloration,[3–11] enabling the broad viewing angles required for their widespread adoption as pigments.[12,13] Furthermore, the inverse structure of the photonic glass has been predicted to reduce the amount of incoherent scattering,[14] extending the range of achievable colors to longer wavelengths and increasing their saturation, however they are even more challenging to produce.[15]

Among the various nanoscale building blocks, block copolymers (BCPs) have been extensively studied as a highly configurable pathway to complex architectures with feature sizes on the order of tens of nanometers; ranging from simple micelles, to lamellae and gyroids.[16,17] However, the discovery of bottlebrush block copolymers (BBCPs), whereby a highly extended backbone is densely grafted with polymer branches, has enabled such architectures to be translated to a larger scale.[18,19] This has unlocked new functionality, particularly in photonics – where BBCPs have been shown to microphase separate into lamellar structures with domain spacings large enough to produce structurally colored photonic crystal films[20–27] and microparticles.[28,29]

Here, we report a new 'controlled micellization' self-assembly mechanism that exploits the ability of amphiphilic BBCPs to behave as 'giant surfactants',[30–33] to produce isotropic porous photonic materials. While amphiphilic BBCPs have been reported to form much larger micelles in water than those obtained with traditional surfactants or BCPs, their sizes usually do not surpass a few tens of nanometers. Here we overcome this limitation by inducing a controlled swelling of reverse BBCP micelles *via* soft confinement within a toluene-in-water microdroplet, resulting in internal aqueous droplets with dimensions comparable to the wavelengths of visible light. Upon subsequent toluene evaporation, the tightly packed nanodroplets self-assemble to template the formation of a highly porous microparticle. The short-range order of the pores within the BBCP scaffold leads to the creation of structural coloration, analogous to inverse photonic glasses. The amount of water available to the micelles can be controlled by temporarily disrupting the microdroplet interface with high shear, allowing the pore size and therefore the color of the resulting structure to be tuned. Such microparticles are the first example of an inverse photonic glass-based pigment, and have several key advantages over their crystalline analogues,[34] including suppressed iridescence and improved color purity, removing the need for refractive index matching or the inclusion of a broadband absorber.

To produce the photonic pigments shown in **Figure 1a**, the 'controlled micellization' of an amphiphilic BBCP within emulsified microdroplets was exploited. (Polynorbornene-*g*-polystyrene)-*b*-(polynorbornene-*g*-polyethylene oxide), P(PS-NB)-*b*-P(PEO-NB), was dissolved in anhydrous toluene and emulsified in water using a rotor-stator homogenizer; see *Methods* for synthesis and characterization. The resultant microemulsion of toluene-in-water droplets was then allowed to evaporate under ambient conditions over *ca*. 30 minutes to yield an aqueous dispersion of vibrantly colored BBCP spherical microparticles ("microspheres"). During this process, any water introduced within the toluene droplet during emulsification is uniformly taken up by the BBCP micelles causing them to swell significantly. Upon final drying, such BBCP-stabilized nanodroplets self-assemble with short-range ordering, giving rise to structural color. By swelling the micelles with an increasing amount of water, the reflected color can be tuned across the visible spectrum (**Figure 1b**), from blue to red, using only a single BBCP ($M_W$ = 290 kDa, $f_{PEO}$ = 50 vol%, DP = 72), as discussed below. This micelle-driven self-assembly mechanism allows for a wide range of colors from a single BBCP; this is in stark contrast to lamellar BBCP photonic systems, which require the precisely controlled synthesis of ultrahigh molecular weight polymers (*ca*. 3000 – 8000 kDa).[28] Unlike the strongly directional character of the reflection from photonic pigments based on ordered photonic crystal structures,[28,35,36] here the coloration originates from the coherent scattering response from inclusions with only short-range order inside the microspheres. This observation is supported by optical microscope investigation under epi-illumination in reflection mode (**Figure 1c**). For a perfectly ordered photonic structure, the reflection is always specular. Therefore in a concentric spherical geometry, the colored reflection spot is localized only at a small region of the microsphere,[28,36] instead here the reflected color is visible across the entire cross-section (**Supporting Figure S1**), confirming a resonant scattering effect as the origin of the coloration.[12]

Cross-sectional analysis of the microspheres by scanning electron microscopy (SEM) confirmed a highly porous internal architecture, which is isotropic throughout their volume (**Figure 2a,b**). The position of the pores exhibits short-range order, i.e. the distance between the pores is well defined only for nearest neighbors (**Figure 2c** and *Methods* for correlation analysis). Comparing microspheres with different colors, as summarized in **Table 1**, we observed that the wall between the pores is consistently in the range 31 – 33 nm, while the average pore diameter and correlation length increase with the reflected wavelength (**Figure 2c** and **Figure 1d** respectively). By emulsifying the BBCP toluene solution in a water-free environment (i.e. toluene-in-perfluorinated oil), lamellae of a similar thickness are observed (**Supporting Figure S2**), indicating that the wall thickness is dependent on the BBCP length (i.e. MW). In contrast, the size of the porous structure is controlled by the amount of water within the volume of the toluene microdroplet, which is nucleated into nanodroplets by the amphiphilic BBCP. This short-range, ordered distribution of pores acts analogously to an inverse photonic glass,[37] where coherent scattering between the low refractive index pores filled with water ($n$ = 1.33) and the BBCP walls (estimated as $n \approx 1.52$) results in a strong reflection at specific wavelengths. Upon removal of water, the microparticle structure is maintained, and the replacement of water in the pores with air ($n$ = 1.00) results in an irreversible blue-shift and a decrease in color saturation (**Supporting Figure S3** and **Supporting Figure S5**). Despite the homogenization process resulting in a polydisperse distribution of microdroplets, the microsphere dimensions determine the intensity but not the wavelength of the reflected color (**Supporting Figure S4**). This is in contrast with direct photonic glass structures, where incoherent scattering limits structural coloration only to a limited range of sizes.[4]

The measured reflection spectra reported in **Figure 1d** for individual red, green and blue BBCP microspheres were compared with the results of numerical simulations, as summarized in **Supporting Figure S5**, with structural parameters derived from the SEM cross-sectional images. The structural color of these short-range ordered architectures is determined by coherent scattering and Mie scattering.[4] The coherent scattering depends on the filling fraction, the average distance between the nearest pores, and the average refractive index. In contrast, Mie scattering is determined by the refractive index contrast of the system, and the average refractive index and diameter of the pores. In the inverse structure, the coherent scattering typically dominates, giving rise to well-resolved spectral peaks.[14] For the sample composed of smaller pores (e.g. blue and green microspheres), only the coherent scattering contributes to the visible optical response as any Mie scattering is mainly at ultraviolet wavelengths. However, when the pore size is large, such that the coherent scattering is in the red, a secondary peak at blue wavelengths caused by Mie resonance is observed.[38,39] In contrast to direct photonic glass and long-range ordered structures, the inverse photonic glass is more tolerant to the variance of pore sizes and correlation length (**Supporting Figure S6**).[14] Such variance results in a broadened ring in the correlation map, which in turns, contributes to a broadened reflection peak. By controlling these features, non-primary colors such as purple, pink, yellow and orange can be successfully achieved (**Supporting Figure S7**).[4,38]

Tuning the optical response of perfectly ordered photonic materials often requires the use of multiple building blocks, for example, by changing the nanoparticle size for polymer opals,[36] or changing the molecular weight in the case of BBCP lamellae.[21] In contrast, by exploiting the controlled swelling of BBCP micelles, here the reflected color can be tuned across the visible spectrum without changing the elemental BBCP building block. This was achieved by varying the speed and/or the duration of the homogenization step, as reported in **Figure 3a** and **Figure 3b** respectively. Interestingly, it was observed that increasing the overall shear experienced during the homogenization step results in both a smaller average microparticle diameter and larger internal pores, with a corresponding red-shift in the optical appearance (a full parameter matrix with microsphere size distributions is included in **Supporting Figure S8**). Furthermore, although the dispersity of the microspheres is significant within a single batch, the small variation in the structural color arising from the internal nano-sized pores, does not correlate with the pigment diameter (**Supporting Figure S1**).

The process leading to the formation of structural color is described schematically in **Figure 3c**. Poly(vinyl alcohol) (PVA) is known to be a emulsion stabilizer,[40] while the symmetric, amphiphilic BBCP can be considered as a giant-surfactant[30–32] that can equally stabilize toluene-water and water-toluene interfaces.[41] As such, upon emulsification, BBCP will assemble at the toluene-water interface of the droplet, with the excess BBCP forming reverse micelles within the volume of the droplet. Any water within the toluene phase will be incorporated into these micelles to reduce interfacial energy, allowing them to grow into nanoscale water droplets that ultimately template the pores observed in **Figure 2b**. The low dispersity in the dimensions of the pores throughout the droplet, which is essential for color production, confirms the controlled swelling of the micelles is a bulk process, rather than from interfacial instabilities.[42,43]

Crucially, given the thickness of the PS block within the BBCP surfactant layer is comparable to the hydrophobic layer in polymersomes,[44] we expect that it can act as a water barrier. Consequently, the ingress of water into the toluene phase by diffusion through the microdroplet interface is very slow. Such intake can however, be significantly enhanced during the high shear achieved during the initial homogenization process, where the increased surface area of the deformed droplets causes the depletion of the interfacial BBCP layer resulting in transient water permeation [45]. Therefore, the overall concentration of water within the toluene phase can be raised by: (i) an increase in homogenizer speed, which will increase the shear during the emulsification process (i.e. raise the rate of water uptake), and (ii) an increase in the homogenization time, which allows for water to be incorporated into the toluene droplets for longer. The greater the amount of water that can be dispersed uniformly *via* chaotic advection within the toluene phase (independent of droplet size), the greater the reverse micelles will swell, resulting in larger pores upon subsequent microparticle formation. This is consistent with the observation that small changes in the homogenization conditions give rise to significant changes in the final color, while increasing the time for diffusion to occur for, by example, slowing the drying time from thirty minutes to three days, results in only a moderate red-shift (**Supporting Figure S9**). Counterintuitively, drying at elevated temperature was found to also red-shift the final color (**Supporting Figure S10**). For example, heating the sample at 40 °C during the toluene evaporation process led to a green dispersion of microspheres rather than the blue dispersion produced under the same emulsion conditions, but dried in the ambient environment (20 °C). This can be explained by increasing the instability of the interfacial BBCP layer upon heating, which can allow for fusion of the internal nanodroplets. In contrast, heating pre-formed pigment microspheres did not result in a color change (tested up to 75 °C), confirming this is not simply an annealing effect.

The emulsion stabilizer was found to influence the final color of the microspheres. Decreasing the PVA concentration from 20 to 0.2 mg/mL resulted in larger droplets with a corresponding blue-shift in the reflected color. This can be explained as a reduction in viscosity and consequently the shear experienced by the BBCP solution (**Supporting Figure S11**). Furthermore, exchanging the PVA with just 0.2 mg/mL of sodium dodecyl sulfate (SDS) results in greater water permeation into the toluene droplet, leading to red-purple microspheres, with an observed decrease in pore size from the surface towards the centre (**Supporting Figure S12**). These observations can be explained by considering that the SDS inhibits the initial formation of the water impermeable BBCP layer, which then allows for greater water diffusion into the droplet during the toluene evaporation process. At higher SDS concentrations, large pore microspheres without short-range correlation were formed.

**Figure 3d** shows a time-lapse of the final stages a microdroplet drying to form a blue microsphere. Starting as a colorless droplet, upon solvent evaporation we observe that the microdroplet appearance turns first blue then to cyan, then it rapidly returns to blue upon complete loss of toluene. While it may be expected that the droplet should initially appear red during the early stages of drying, given that the water nanodroplets gradually pack more closely together within the shrinking toluene microdroplet; its absence instead suggests that the nanodroplets are randomly arranged without any correlation at lower BBCP concentration. When the color appears, the size of the toluene droplet is close to that of the final dried state, suggesting the short-range correlated structure is formed only at this late point in the drying process. Therefore, removal of the remaining toluene first leads to an increase in average refractive index ($n_{\text{BBCP}} > n_{\text{tol}}$), giving rise to the cyan color, while the final compression of the polystyrene domains, upon complete loss of solvent, causes the subsequent blue-shift. Interestingly, this late-stage self-assembly mechanism is in stark contrast to interfacial-templated approaches,[28] and is significant because it enables a more rapid and robust route to produce photonic pigments.

To test the performance of the fabricated photonic materials as pigments, in terms of iridescence and off-specular color response, we investigated the macroscopic visual appearance of the microspheres dispersed in water and as a dried film. Photographs of green microspheres dispersed in water inside a flat glass capillary were collected, while it was rotated relative to a fixed illumination and observation angle. As shown in **Figure 4a**, a consistent green appearance was observed throughout the angular range. The lack of iridescence upon rotation under direct illumination at a fixed angle confirms the isotropy of the spherical particles. Note that although the structure is isotropic, under direct illumination multiple Bragg conditions can be satisfied simultaneously.[46] As such, when observing away from the illumination direction, shorter wavelengths will be scattered, leading to iridescence (**Supporting Figure S13**). However, under diffuse illumination, the dominant reflection is attributed by the light that comes from the observer's viewpoint, making the color consistent regardless of the observer's position. Finally, a dispersion of red microspheres was dried in air on a glass substrate to yield a green monolayer film (see **Supporting Figure S3** for discussion of the blue-shift upon drying in air). This film similarly showed a consistent green appearance across the same angular range. Such results are significant, as photonic pigments based on ordered structures typically require either refractive index matching or the inclusion of a broadband absorber to avoid scattering-induced effects.

In summary, we demonstrate that the controlled swelling of amphiphilic BBCP micelles within an emulsified droplet is a convenient process to directly fabricate angular-independent photonic pigments covering the full visible spectrum. Such optical properties are the result of a short-range distribution of pores leading to coherent scattering effects. We believe that this is a particularly interesting route to produce photonic pigments, not only for the optimal optical performance, with vibrant color visible even at extremely low concentrations and under natural illumination, but also for its potential scalability. The microspheres were prepared in a single processing step, with one polymeric component, with the final structural color determined only by the intensity of the homogenization process. This fabrication method is advantageous over existing methods, such as the swelling of linear BCP micelles with homopolymer,[47] as it directly allows access to the large pore sizes required for photonics. While a greater refractive index contrast could be achieved with porous inorganic structures, potentially allowing for a stronger optical response, such systems typically require multiple fabrication steps, including self-assembly, infiltration with an inorganic precursor, and aggressive removal of the nanoparticle template. In contrast to methods explored to produce photonic crystal pigments, here the self-assembly was an isotropic process within the volume of the droplet, rather than templated by the droplet interface, resulting in a comparatively rapid fabrication time, high tolerance to structural defects and color-independence from the microparticle dimensions (avoiding the need for e.g. microfluidics). As such, these scalable photonic pigments offer the broad viewing angles and fade-resistance required for applications ranging from colorants in automotive paints and exterior coatings to use as pixels in reflective displays.

## Experimental Section

*Synthesis of (Polynorbornene-g-polystyrene)-b-(polynorbornene-g-polyethylene oxide) BBCP*: The BBCP was synthesized according to the reported procedure.[48] In a typical experiment, 100 mg of PS-NB ($M_n$ = 4 kDa, PDI = 1.14) and PEO-NB ($M_n$ = 4 kDa, PDI = 1.05) macromonomers (MMs) were added to separate Schlenk flasks followed by the desired amount of anhydrous DCM. The concentration of the PS-NB MM was controlled from 0.05 – 0.1 M.

The resulting solutions were degassed with three cycles of freeze-pump-thaw before sequential polymerization. At room temperature, the polymerization of PS-NB was initiated by adding the desired amount of third-generation Grubbs catalyst solution in DCM. After the first MM, PS-NB, reacted for 60 min, the solution of the second MM, PEO-NB, was injected into the reaction mixture. This solution was stirred for an additional 2-3 hours before being quenched with ethyl vinyl ether. Conversions of PS-NB and PEO-NB were all *ca.* 100% based on $^1$H-NMR spectra. The volume fraction of the PEO block was controlled at 50 vol% as determined using $^1$H-NMR spectroscopy. The weight-average molecular weight ($M_w$) of the BBCP is 290 kDa (PDI = 1.06), as determined by gel permeation chromatography equipped with multi-angle laser light scattering detector and a refractive index detector.

*BBCP microsphere formation:* The P(PS-NB)-*b*-P(PEO-NB) BBCP (20 mg) was added to a vial followed by toluene (1.0 mL) to afford a transparent and colorless stock solution (20 mg/mL). For the aqueous phase, poly(vinyl alcohol) (PVA, 200 mg) was employed as a stabilizer and as such, was dissolved with heating into Milli-Q water (10 mL) to form a clear solution (20 mg/mL). Polydisperse microdroplets were generated using an IKA T25 digital Ultra-Turrax homogenizer. To prepare a dispersion of microspheres, the homogenizer tip was located at the base of a glass vial (2 Dram) containing the aqueous PVA solution (5 mL) and rotated at 6,000 – 14,000 rpm. Upon injecting the BBCP solution (10 μL) adjacent to the spinning homogenizer impeller, an emulsion of toluene-in-water microdroplets was immediately formed. The emulsification process is assumed to be a lossless process, i.e. all BBCP solution is converted into microdroplets. After a controlled period of time (2 – 120 s) the homogenizer was stopped and the formed microemulsion floated to the air-water interface, upon which toluene evaporation over ~30 minutes resulted in colored microspheres. Final loss of toluene from the microspheres caused them to sediment and for the color to further evolve. The produced porous, photonic microspheres were washed with Milli-Q water to remove residual PVA surfactant and dried under nitrogen flow to yield a dispersible powder. To extend the drying time a lid was fitted to the vial, while heating in an oven was used to tune the color post-homogenization. When stored as a dispersion in water over a period of twelve months, such pigments showed no evidence of degradation or any apparent change in their optical appearance.

*BBCP microsphere characterization:* Optical microscopy and micro-spectroscopy were performed on a customized Zeiss Axio Scope A1 microscope fitted with a CCD camera (Eye IDS, UI-3580LE-C-HQ, calibrated with a white diffuser) using a halogen lamp (Zeiss HAL100) as a light source in Koehler illumination. To perform micro-spectroscopy, the microscope was coupled to a spectrometer (Avantes, AvaSpec-HS2048) using an optical fiber (Avantes, FC-UV100-2-SR, 50 µm core size). The reflectance spectra were normalized against a white diffuser (Labsphere SRS-99-010). The microscope was set-up so that the numerical aperture (NA) in illumination is limited by the NA of the objective used. The BBCP microspheres dispersed in water were analyzed using a water immersion objective (Zeiss, W N-Achroplan, 63x, NA 0.9) and the dried BBCP microspheres were analyzed using a 50x objective (Zeiss, LD EC Epiplan-Neofluar, NA 0.55). Photographs were recorded with a digital SLR camera (Nikon D3200) at a pigment concentration of 0.03 wt% in water and against an absorbing background. The effect of the background or the microsphere concentration upon the vibrancy of the pigment dispersion is illustrated in **Supporting Figure S14**. It should be noted that there is sufficient size dispersity for microparticles made by different conditions and with consequently different colors to overlap in diameter (see **Supporting Figure S8**). As such, to aid in the understanding of the optical appearance and to link it back to the trend in the underlying porous structure, microparticles of a similar size were selected for Figures 1 and 2.

*Scanning electron microscopy (SEM):* Images were collected with a Mira3 system (TESCAN) operated at 5 kV and a working distance of 6 mm. The samples were mounted on aluminum stubs using conductive carbon tape and coated with 10 nm thick layer of Pt with a sputter coater (Quorum Q150T ES). The samples were fractured mechanically by a razor blade to expose the cross-section.

*Correlation analysis:* The degree of order of the porous internal architecture of the microspheres was quantified with the structure factor – i.e., the Fourier transform of the centre of the pores.[49,50] The structure factor allows for direct estimate of the average distance between pores and their spatial organization.

*Simulation:* Disordered two dimensional structures with short-range correlation were generated using a recently developed inverse design algorithm, as discussed in detail in Jacucci *et al*.[49] This algorithm allows for the simulation of inverse photonic glasses whose single-particle (size distribution) and structural (correlation length) properties can be controlled independently. Input parameters were derived from correlation analysis of the SEM cross-section images in Figure 2, as discussed in the Supporting Information. Numerical simulations of the optical response of the generated structures were performed in Lumerical (Lumerical Inc.), a software using the finite difference time domain (FDTD) method.

## Acknowledgements

This work was supported by the European Research Council [ERC-2014-STG H2020 639088; ERC-2017-POC 790518], the BBSRC David Phillips Fellowship [BB/K014617/1], the EPSRC

**Tables**

|       | Homogenizer speed [rpm] | Homogenization time [s] | Particle diameter [μm] | Wall thickness [nm] | Estimated pore size [nm] |
|-------|-------------------------|-------------------------|------------------------|---------------------|--------------------------|
| Blue  | 6000                    | 5                       | 29.0 ± 7.8             | 31.3 ± 3.3          | 133 ± 22                 |
| Green | 8000                    | 15                      | 19.9 ± 3.8             | 32.9 ± 2.3          | 150 ± 52                 |
| Red   | 8000                    | 60                      | 16.3 ± 3.0             | 33.3 ± 2.8          | 193 ± 43                 |

**Table 1.** Summary of the fabrication conditions and physical parameters of the red, green, and blue pigments reported in Figure 1 (optical appearance) and Figure 2 (structural characterization). The particle diameter and wall thickness were measured directly from SEM images. The pore size was estimated from the SEM cross section using the 'minimal circle fit for contour' function in *OpenCV*. Note, the large variance is not attributed to the dispersity of the pores themselves, but to the limitations in reconstructing the 3D porous architecture from a cross-sectional SEM, which is subject to variation depending on the cut angle through each pore and any asphericity arising from the tight packing upon final drying of the swollen micelles.

**Figures**

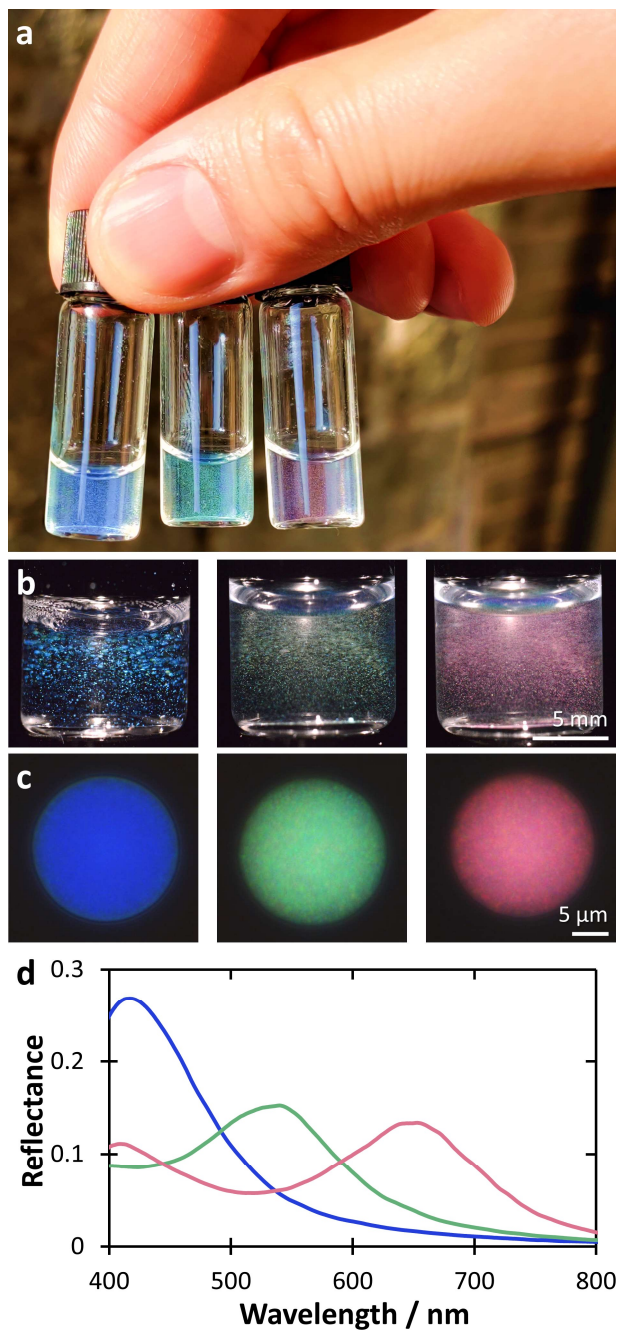

**Figure 1. (a)** Aqueous dispersions of blue, green, and red BBCP photonic pigments, illuminated under natural sunlight. **(b)** Photographs showing the same 0.03 wt% dispersions under direct illumination, and **(c)** micrographs of individual microspheres collected in epi-illumination. **(d)** Corresponding reflectance spectra for the microspheres shown in (c), with the intensity measured relative to a white Lambertian diffuser.

19**Figures**

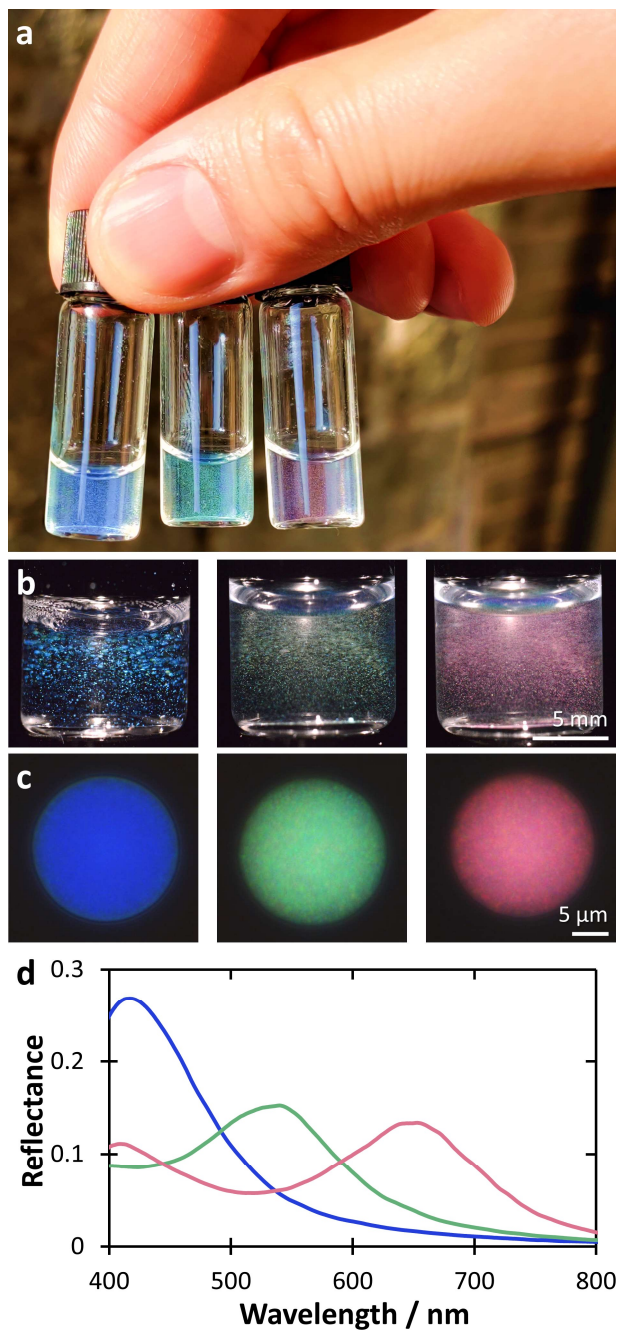

**Figure 1. (a)** Aqueous dispersions of blue, green, and red BBCP photonic pigments, illuminated under natural sunlight. **(b)** Photographs showing the same 0.03 wt% dispersions under direct illumination, and **(c)** micrographs of individual microspheres collected in epi-illumination. **(d)** Corresponding reflectance spectra for the microspheres shown in (c), with the intensity measured relative to a white Lambertian diffuser.

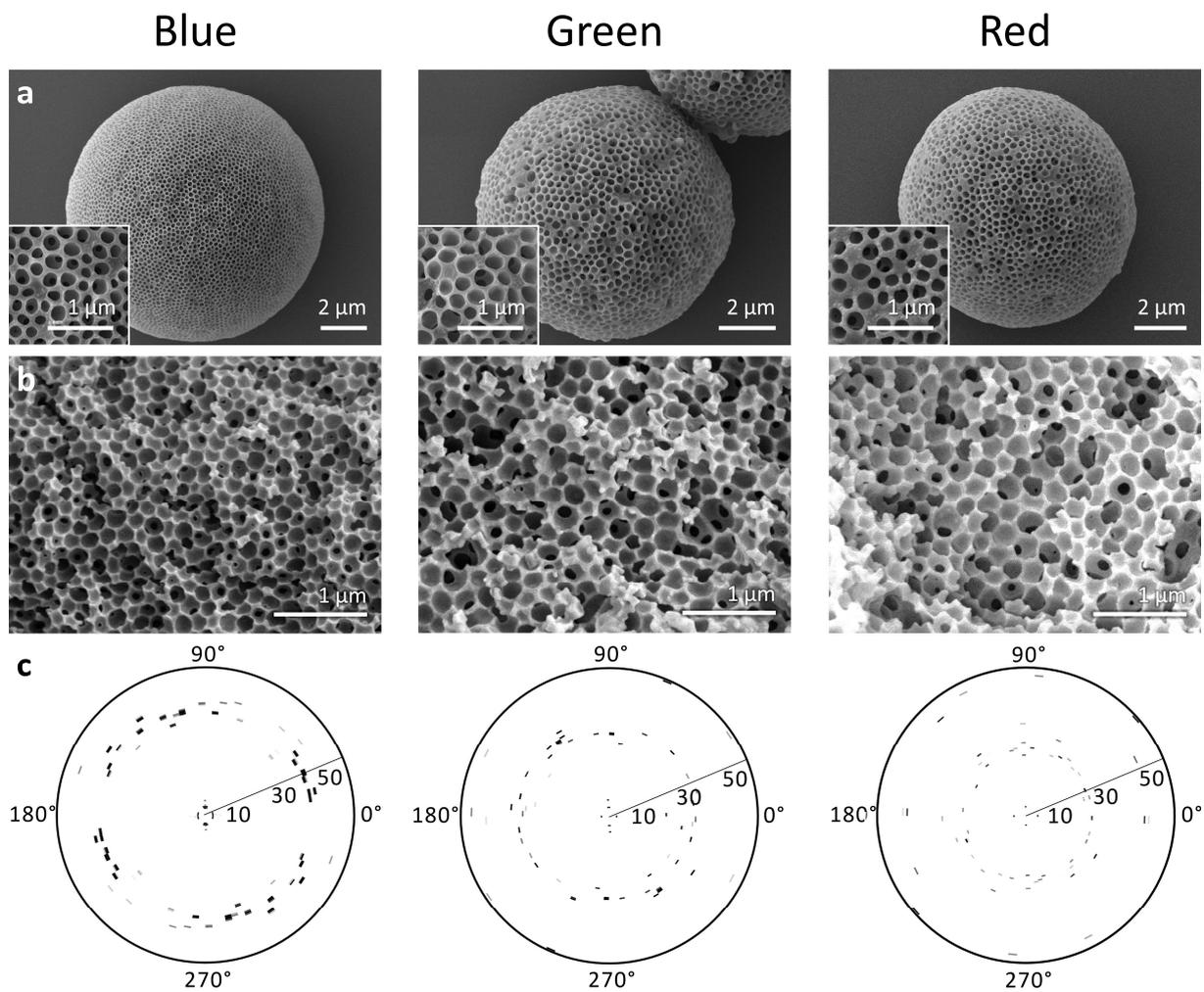

**Figure 2.** Scanning electron micrographs showing **(a)** the porous surface of the BBCP microspheres, and **(b)** cross-section of the internal architecture, exposing the highly porous structure with short-range order. **(c)** Polar (wavevector-angle) plots showing the structural correlation of the micrographs in (b).

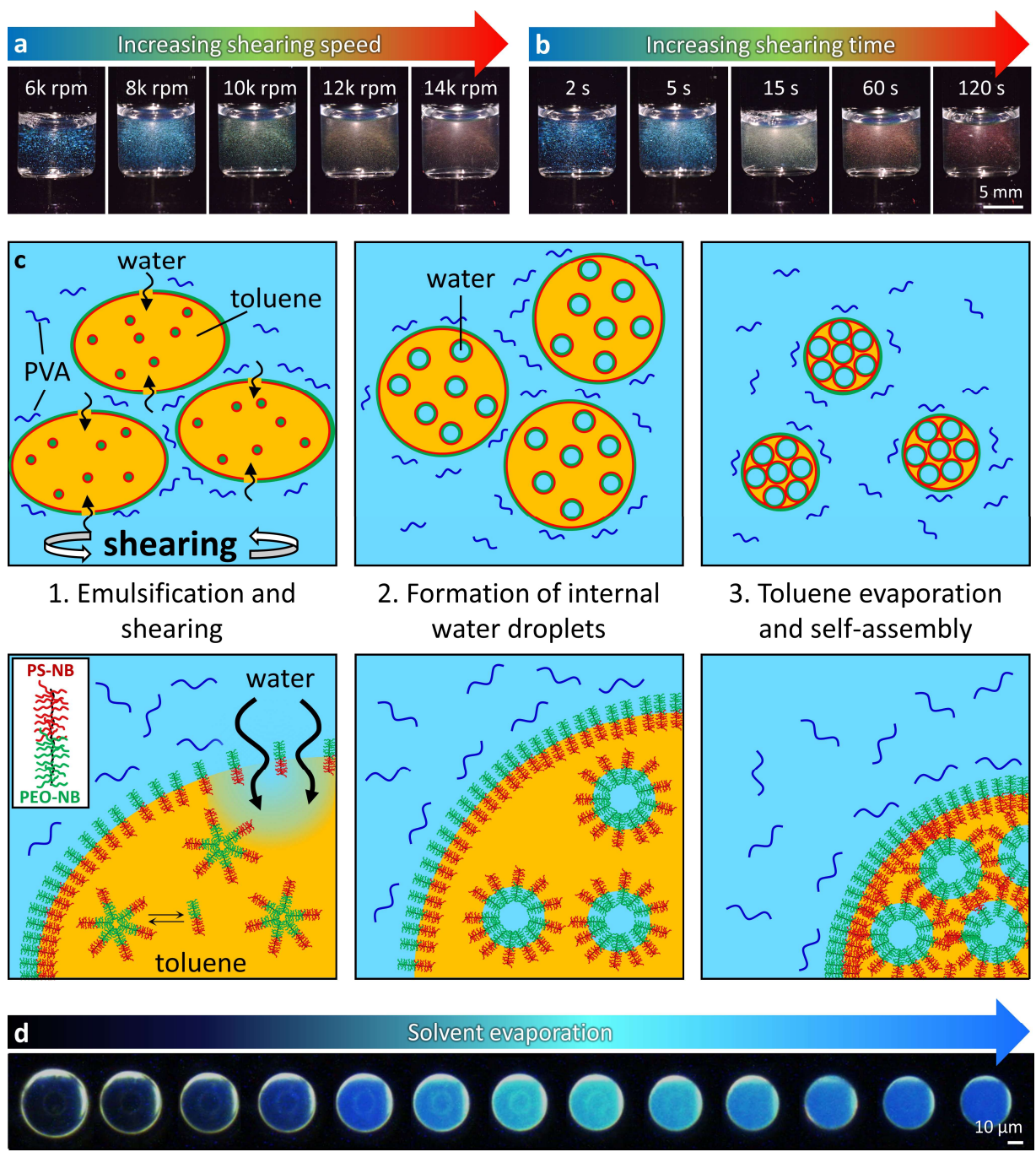

**Figure 3.** The reflected color from the porous BBCP microspheres can be tuned by either **(a)** the homogenization speed (5 s: 6000 – 14000 rpm) or **(b)** homogenization time (8000 rpm: 2 – 120 s). **(c)** Schematic showing the formation of porous, 'inverse photonic glass' microspheres from an evaporating toluene-in-water emulsion containing the amphiphilic P(PS-NB)-*b*-P(PEO-NB) BBCP. **(d)** Time-lapse micrograph series recorded in darkfield, showing the sequence of color changes upon the final stages of drying to form a blue microsphere.

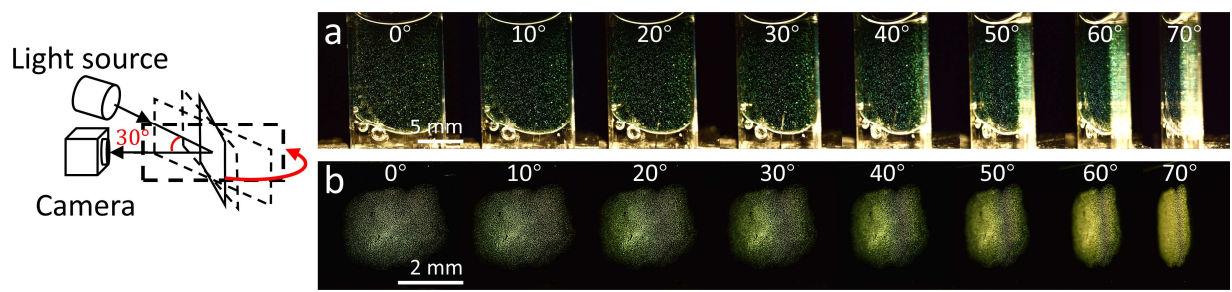

**Figure 4.** Angular dependent macroscopic photographs of the photonic pigment as: **(a)** an aqueous dispersion of green microspheres inside a flat glass capillary, showing the consistent green color upon rotation of the capillary; **(b)** dried green microspheres on a glass slide, showing the visible green color without any additional refractive index matching, over the same range of angles as in (a).

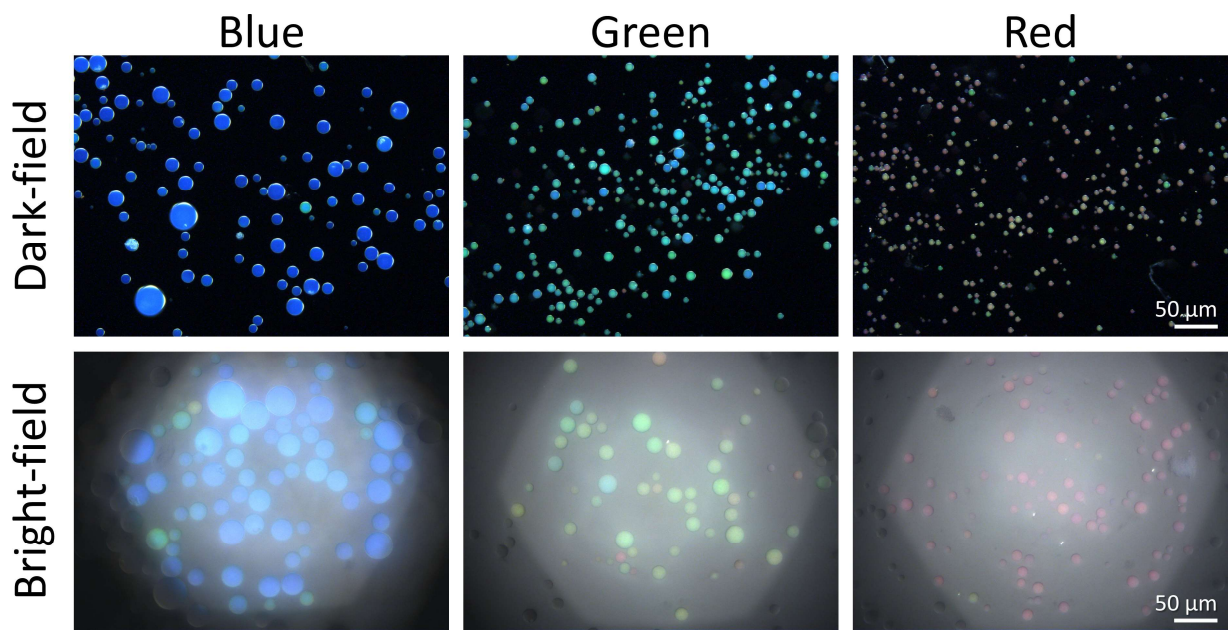

**Figure S1 | Optical characterization of BBCP microspheres**. Comparison between dark-field and bright-field reflection micrographs of the blue, green and red BBCP microspheres, showing that the coloration originates form a coherent scattering response, rather than the strongly directional character observed for ordered photonic crystal structures. The blue-shift when illuminated under dark-field is consistent with Bragg's law and is due to the increased illumination angle. The images were taken using a customized Zeiss Axio scope A1 inverted microscope. The sample, an aqueous drop of the BBCP dispersion, was on the top surface of a microscope slide and imaged from below (i.e. through the glass). The dark-field images were taken with a Zeiss EC Epiplan-Apochromat objective (10x NA 0.3) and the bright-field images were taken with a Zeiss EC Epiplan-Apochromat objective (20x NA 0.6).

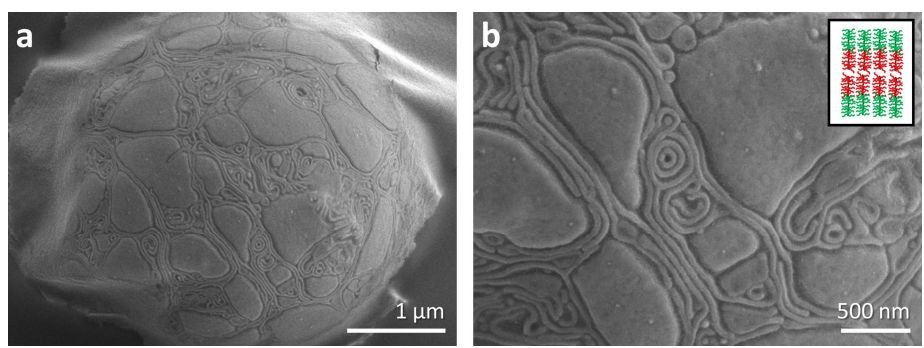

**Figure S2 | Emulsification of the amphiphilic BBCP in a water-free environment.** Emulsification of the BBCP toluene solution (20 mg/mL) in perfluorinated oil (FC-40 with 2.0 wt% XL-01-171 surfactant, Sphere Fluidics) resulted in solid microspheres that do not display structural color. Scanning electron microscopy revealed worm-like lamellae across the surface of the microsphere, corresponding to self-assembled BBCP with limited long-range ordering. The periodicity of the lamellae is comparable to the wall thickness between the pores in the toluene-in-water system, suggesting that the wall thickness is dependent on the BBCP length (i.e. MW).

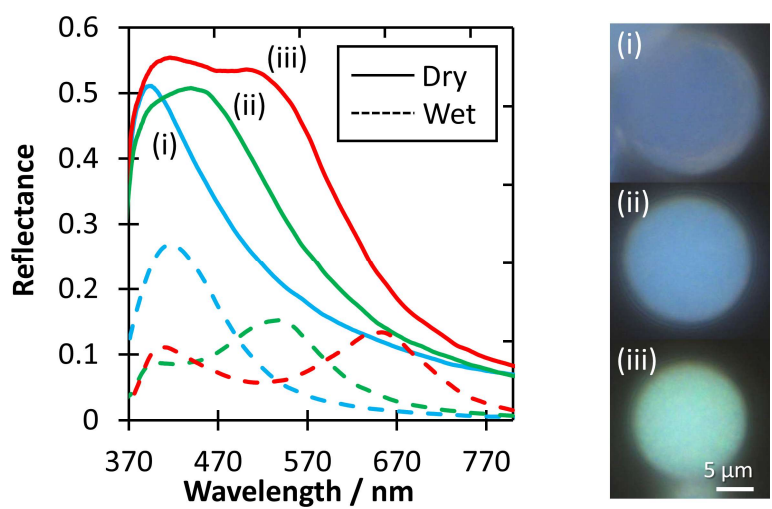

**Figure S3 | The change in the optical appearance of BBCP microspheres upon drying in air.** Upon loss of water ($n = 1.33$) from the pores of the (i) blue, (ii) green and (iii) red microspheres and replacement with air ($n = 1.00$), a shift towards blue wavelengths with a decrease in color saturation is observed.

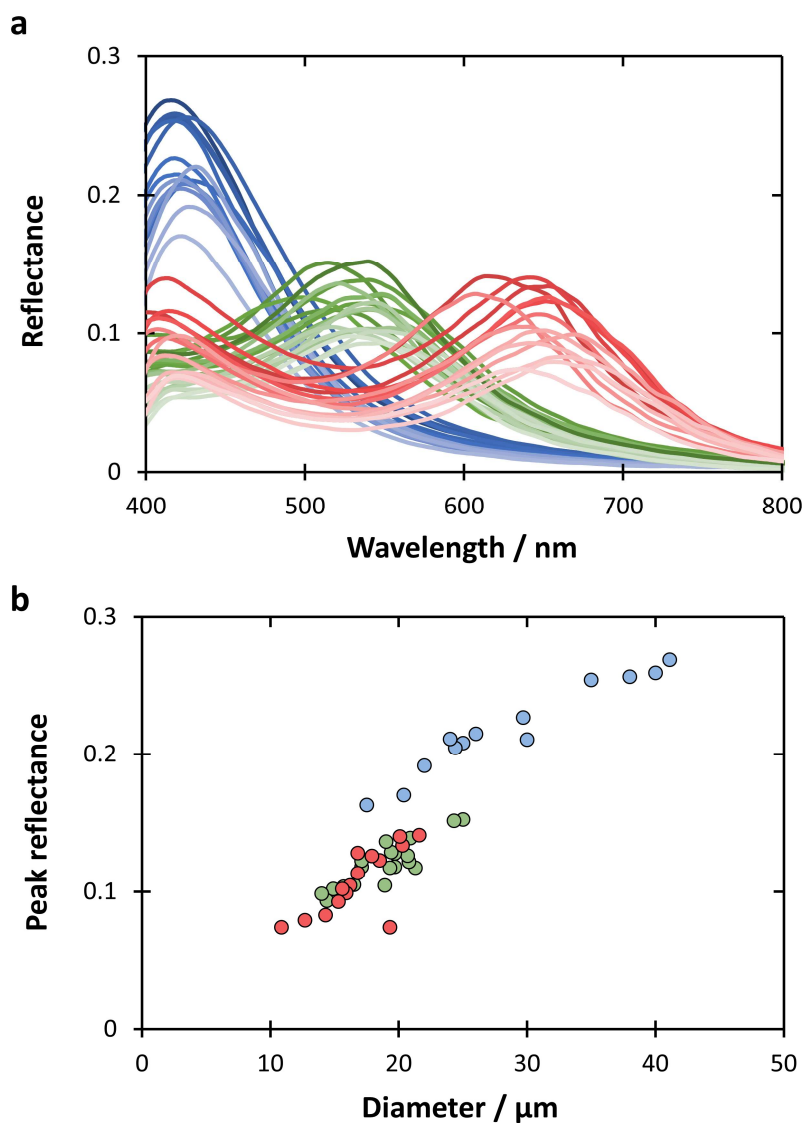

**Figure S4 | The variance in peak reflectance for blue, green, and red microspheres as a function of size. (a)** The final microsphere diameter is typically smaller for more intense homogenization conditions (i.e. red microspheres are on average smaller), but the color does not vary within each polydisperse population as a function of size. **(b)** The intensity of the reflected color follows a linear correlation with the diameter of the microsphere, and this trend is general, irrespective of the peak wavelength of the microsphere.

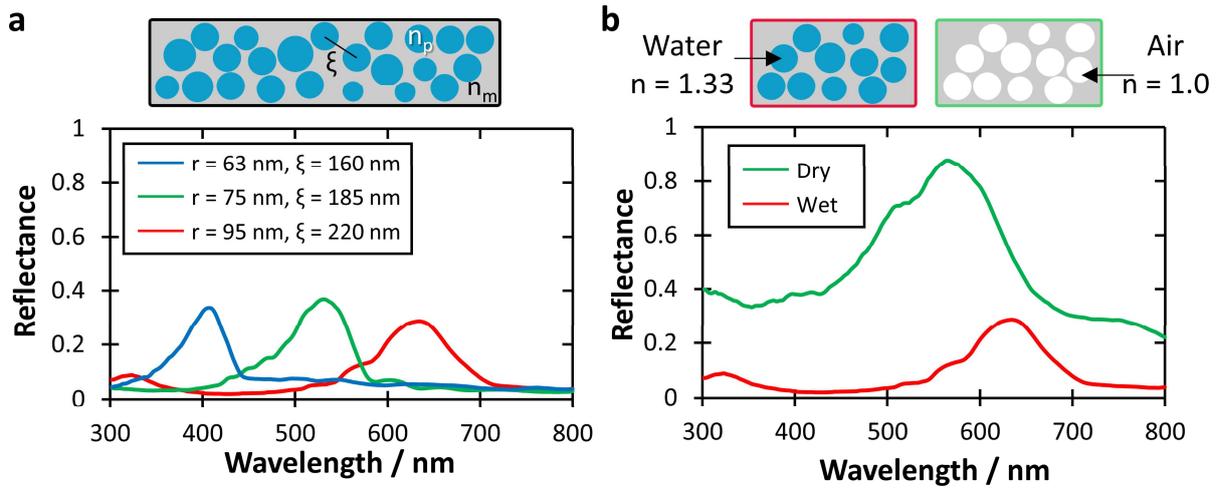

**Figure S5 | Simulated optical response for inverse structures with short-range order (I).** The simulation parameters were varied as a function of **(a)** the correlation distance and **(b)** the refractive index of the inclusions. The schematics illustrate the simulation parameters. All simulations assume the filling fraction of the inclusions is 40% and an overall system thickness of 5 μm. The distribution of the size of the inclusions and the value of the correlation distance were extracted from the SEM micrographs in Figure 2 of the article.

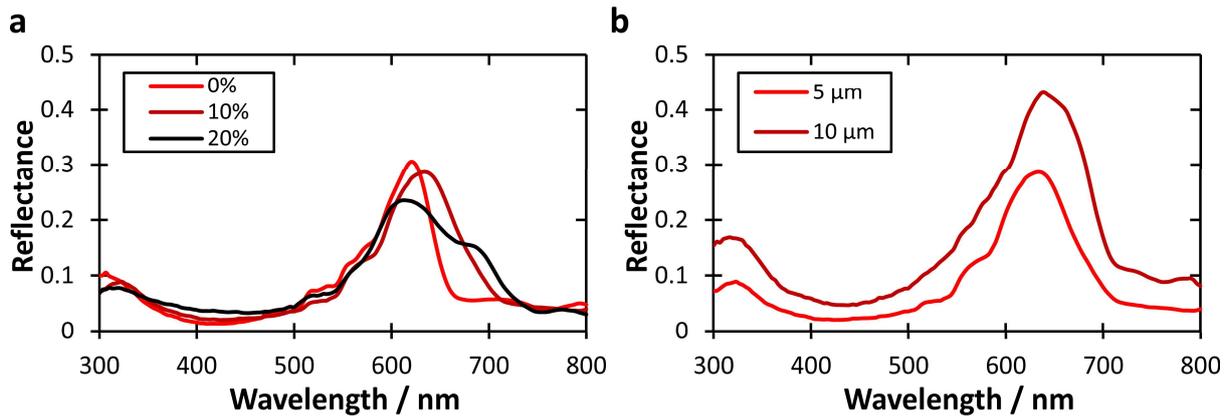

**Figure S6 | Simulated optical response for inverse structures with short-range order (II).** The simulation parameters were varied as a function of **(a)** the polydispersity of the inclusions and **(b)** the overall size of the system. All simulations assume the filling fraction of the inclusions is 40%.

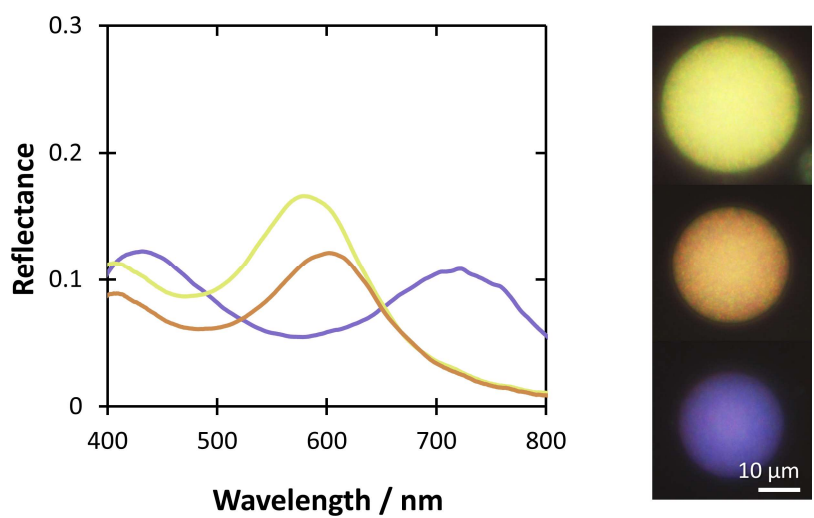

**Figure S7 | The production of BBCP microspheres with non-primary colors.** The variance in correlation between the pores contributes to a broadened reflection peak that can produce non-primary colors such as yellow and orange, while purple can be produced by exploiting the presence of a secondary peak at blue wavelengths for red-infrared microspheres.

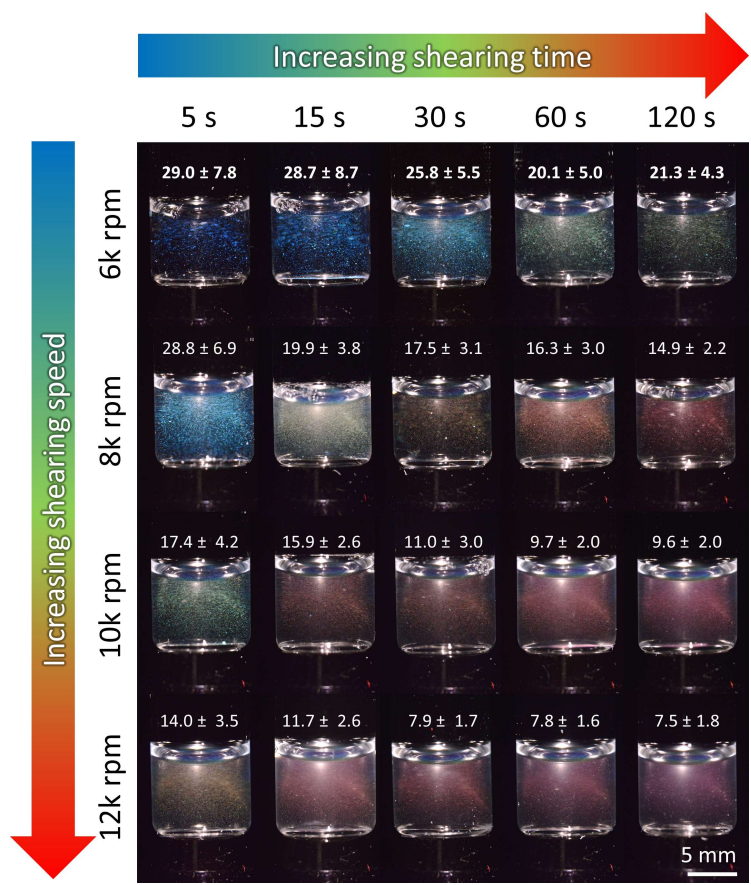

**Figure S8 | Visual summary of the wide spectrum of colors that can be achieved by changing the homogenization conditions.** Control over the homogenization speed and time allows for the color of the microspheres to be straightforwardly and precisely tuned from blue to red. The average size of the microsphere dispersion is indicated above each photograph and shows a general trend that increasing the homogenization intensity leads to smaller, red-shifted microspheres.

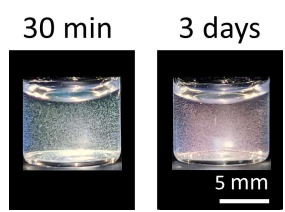

**Figure S9 | The effect of drying time on the final color of the BBCP microspheres.** By drying an emulsion with the vial cap on, the drying time could be significantly delayed from *ca*. 30 minutes to 3 days. This results in a red-shift of the final color of the microsphere dispersion.

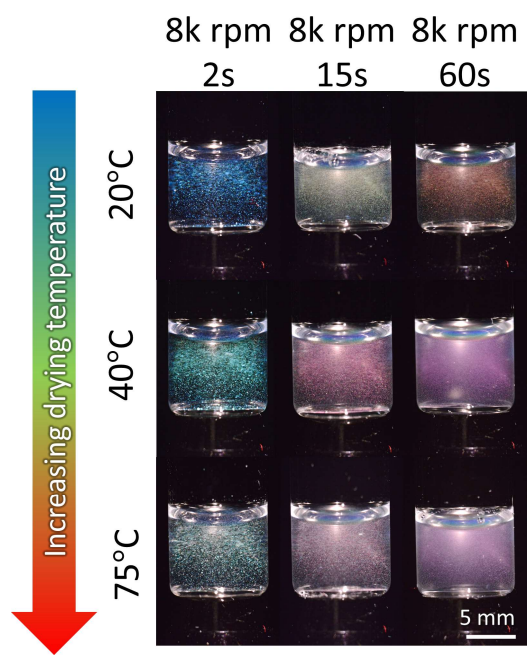

**Figure S10 | Drying the BBCP toluene-in-water microemulsion at an elevated temperature.** Oven drying results in a red-shift of the final color (40 °C), when compared to the same emulsion dried under ambient conditions (20 °C). Drying at a higher temperature (75 °C) results in a further red-shift (albeit small), but with loss of color purity, attributed to the higher rate of toluene loss resulting in a less well-correlated pore structure.

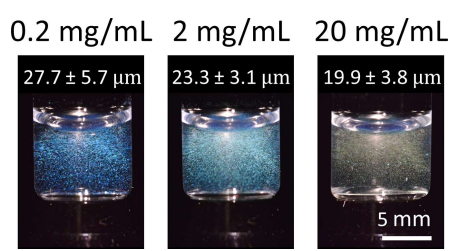

**Figure S11 | The effect of the PVA surfactant concentration on the final color of the BBCP microspheres.** Upon increasing the PVA concentration from 0.2 – 20 mg/mL, there is a red-shift in the final color in combination with a decrease in the average size of the microspheres, as indicated above each photograph. This effect is attributed to a rise in viscosity of the continuous phase, which increases the shear experienced by the BBCP solution during homogenization.

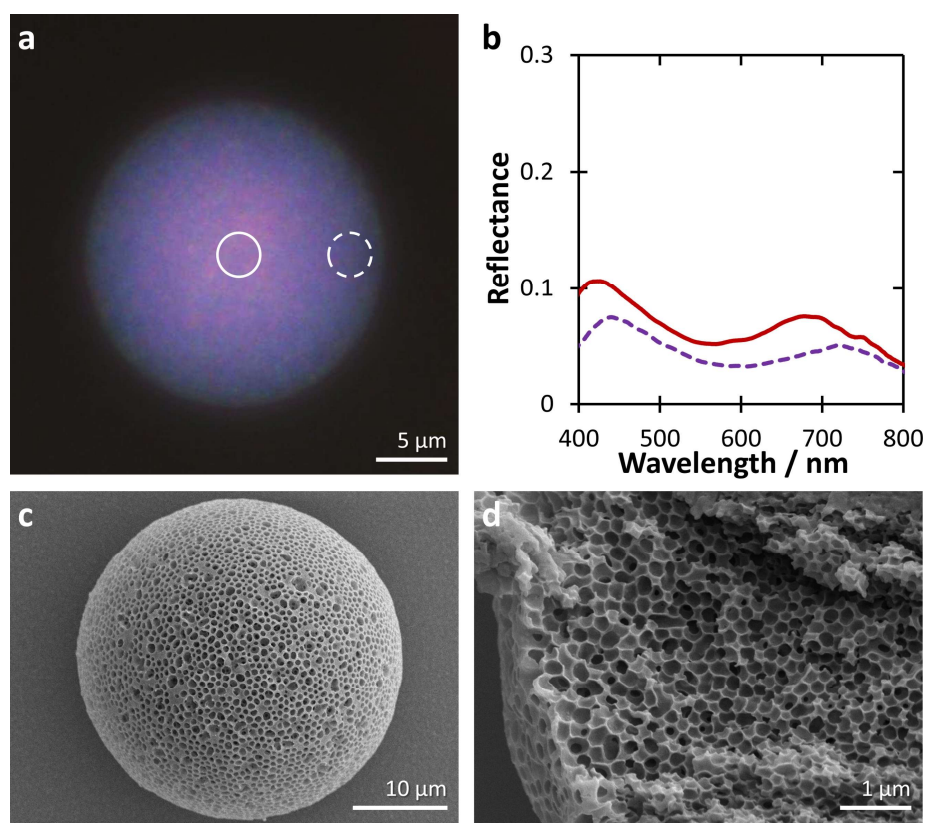

**Figure S12 | BBCP microspheres prepared using sodium dodecyl sulfate (SDS) as a surfactant. (a)** Exchanging the surfactant from polyvinyl alcohol (PVA, 20 mg/mL) to SDS (0.2 mg/mL) results in significantly red-shifted microspheres (with the peak at ~420 nm wavelength leading to an overall purple appearance). Additionally, the color is no longer uniform across the particle, instead the color shifts from red at the core to purple at the surface. **(b)** Micro-spectroscopy at the two regions marked by circles in (a) confirmed these observations. **(c,d)** Scanning electron microscopy confirmed that the pores are not only bigger than typically seen with PVA, but also that the pores near the surface are larger. This can be explained by the presence of SDS at the droplet interface inhibiting the initial formation of a water impermeable BBCP layer and thus enabling a greater rate of water diffusion into the toluene droplet. As this continues to occur after the initial shear during homogenization, the water is no longer uniformly distributed throughout the droplet, resulting in the pores at the interface swelling preferentially.

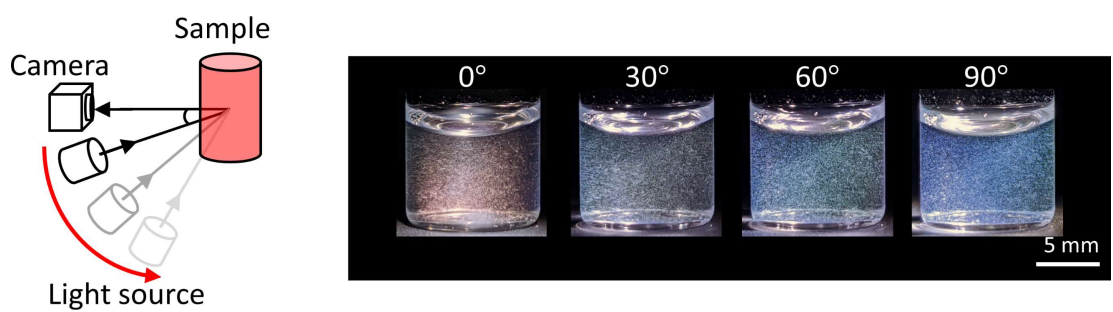

**Figure S13 | Angular response from BBCP microspheres when illuminated under direct illumination.** Photographs of an aqueous dispersion of red microspheres in a vial under direct illumination. Upon increasing the angle between the light source and the observer a strong blue-shift is observed. This iridescence is not observed when the vial is under diffused illumination because the dominant reflection always originates from the observer's position.

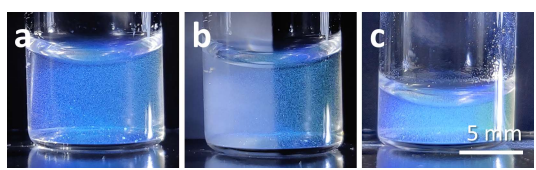

**Figure S14 | Assessing the optical performance of the BBCP microspheres. (a)** Photograph of an aqueous dispersion of red microspheres (0.03 wt%) in a vial under direct illumination at 90° to the observer, resulting in a strongly blue-shifted appearance. **(b)** Although P(PS-NB)-*b*-P(PEO-NB) is colorless (i.e. not absorbing), the selective reflection of a narrow bandgap of wavelengths in the visible spectrum by the nanostructured BBCP microsphere results in visible color. All other wavelengths are transmitted or scattered at the microparticle interface. As such the contrast of the BBCP pigment is reduced if placed in front of a scattering white background (*left*) but is enhanced by an absorbing black background (*right*). **(c)** Doubling the concentration of the pigment dispersion to 0.06 wt% results in a more intensely colored dispersion.

# Appendix 1: Further discussion of the optical simulation

To model the optical response of the BBCP microspheres, we simulated the optical properties using a reconstructed porous architecture (see Methods section in the article). This 2D slab model disregards the curved pigment interface and any specific illumination conditions under the microscope. As inputs, the center-to-center (correlation) distance and the perceived filling fraction were extracted from analysis of the SEM images in Figure 2b of the article, with the average pore diameter estimated from the center-to-center distance minus a BBCP-wall thickness of 32 nm. It is believed that estimating the pore size in this way is more accurate than simply measuring the diameters of the pores visible in the SEM cross-section, as it is unlikely to cut through the equator of each pore and would also misrepresent any dispersity in pore size. The average refractive index of the BBCP was estimated from the reported values for the two constituent side chains and the volume fraction of 50 vol% PEO by $^1$H-NMR spectroscopy, whereas the pores were considered as either pure water or air. By constructing this simplified model, parameters such as the pore size dispersity and pigment thickness could be varied allowing their impact on the optical response to be explored.

However, it should be noted that the simulated and experimental spectra, Figure S5 and Figure 1d of the article respectively, show a small discrepancy in terms of the position of the peaks (i.e. blue-shifted in the simulations) and also in the weaker intensity of the secondary peak in the simulated red structure. This divergence can arise from:

(i) *Experimental uncertainty*. For example, if the porous microparticle has contracted between the optical analysis in solution and the SEM cross-sectional analysis under high vacuum, a blue-shift would be observed in the simulated spectra due to undermeasuring of the internal dimensions of the solvated porous structure.

(ii) *Simulation assumptions*. The simulation represents a 2D projection of the 3D porous BBCP architecture, however it is rebuilt from the analysis of a 2D image of the microparticle cross-section. As such, assumptions were made to extract the required parameters. For example, the center-to-center distance is likely to be an underestimate as the projection of the pore centers out of the cross-sectional plane are not accounted for. This would be expected to result in a smaller estimated pore diameter leading to the blue-shift of the peaks in the simulated spectra.